\documentclass{aa}  
\usepackage{textcomp}
\usepackage[utf8]{inputenc}
\usepackage{lscape}
\usepackage{lastpage}
\usepackage{amsmath,amssymb}
\usepackage{mathrsfs}	
\usepackage{multicol}

\usepackage{graphicx}
\usepackage{txfonts}
\usepackage{hyperref}
\hypersetup{
    colorlinks   = true, 
    urlcolor     = cyan, 
    linkcolor    = blue, 
    citecolor   = blue 
}
\usepackage{placeins}
\usepackage{xcolor}
\usepackage{caption}

\graphicspath{plots}

{\newif\ifnotend
\notendtrue
\def\veclist{ABCDEFGHIJKLMNOPQRSTUVWXYZabcdefghijklmnopqrstuvwxyz.}
\def\top#1#2.{#1}
\def\tail#1#2.{#2.}
\loop\expandafter\xdef\csname bb\expandafter\top\veclist\endcsname%
{{\noexpand\bf\expandafter\top\veclist}}
\edef\veclist{\expandafter\tail\veclist}
\if\veclist.\notendfalse\fi\ifnotend\repeat}
\def\d{{\rm d}}

\newcommand{\kpc}   {\,{\rm kpc}}
\newcommand{\pc}    {\,{\rm pc}}

\newcommand{\Msun}  {\,M_{\odot}}

\newcommand{\kms}   {\,{\rm km\,s^{-1}}}
\newcommand{\asec}  {\,{\rm arcsec}}
\newcommand{\amin}  {\,{\rm arcmin}}

\newcommand{\fJ}    {f({\boldsymbol J})}
\newcommand{\fJi}   {f_i({\boldsymbol J})}
\newcommand{\fJst}  {f_\star({\boldsymbol J})}

\newcommand{\hJi}   {h_i({\boldsymbol J})}

\newcommand{\gJi}   {g_i({\boldsymbol J})}

\newcommand{\fni}   {f_i}
\newcommand{\Mi}    {M_i}
\newcommand{\Mst}   {M_\star}
\newcommand{\Mdm}   {M_{\rm dm}}
\newcommand{\MBH}   {M_{\rm BH}}

\newcommand{\Ji}    {J_i}
\newcommand{\Jst}   {J_{\star}}
\newcommand{\Jdm}   {J_{\rm dm}}

\newcommand{\Jti}    {J_{{\rm t}, i}}

\newcommand{\Jtdm}   {J_{\rm t, dm}}

\newcommand{\Jci}    {J_{{\rm c}, i}}
\newcommand{\Jcst}   {J_{{\rm c},\star}}
\newcommand{\Jcdm}   {J_{\rm c, dm}}

\newcommand{\alphai}    {\alpha_{i}}
\newcommand{\alphast}   {\alpha_\star}
\newcommand{\alphadm}   {\alpha_{\rm dm}}

\newcommand{\etai}   {\eta_{i}}
\newcommand{\etast}  {\eta_\star}
\newcommand{\etadm}  {\eta_{\rm dm}}

\newcommand{\Bi}   {B_{i}}
\newcommand{\Bst}  {B_\star}
\newcommand{\Bdm}  {B_{\rm dm}}

\newcommand{\Gammai}   {\Gamma_{i}}
\newcommand{\Gammast}  {\Gamma_\star}
\newcommand{\Gammadm}  {\Gamma_{\rm dm}}

\newcommand{\hri}  {h_{r,i}}

\newcommand{\hzi}  {h_{z,i}}
\newcommand{\hzst} {h_{z,\star}}
\newcommand{\hzdm} {h_{z,{\rm dm}}}

\newcommand{\gri}  {g_{r,i}}

\newcommand{\gzi}  {g_{z,i}}
\newcommand{\gzst} {g_{z,\star}}
\newcommand{\gzdm} {g_{z,{\rm dm}}}

\newcommand{\Jr}    {J_r}
\newcommand{\Jphi}  {J_{\phi}}
\newcommand{\Jz}    {J_z}

\newcommand{\JJ}    {\boldsymbol{J}}

\newcommand{\Phitot}{\Phi_{\rm tot}}
\newcommand{\PhiBH} {\Phi_{\rm BH}}
\newcommand{\Phist} {\Phi_\star}
\newcommand{\Phidm} {\Phi_{\rm dm}}
\newcommand{\Phii}  {\Phi_i}

\newcommand{\LL}    {\mathcal{L}}
\newcommand{\LLd}   {\mathcal{L}_{\Sigma_\star}}
\newcommand{\LLs}   {\mathcal{L}_{\sigma_{\rm los}}}
\newcommand{\LLvd}  {\mathcal{L}_{\mathscr{L}}}
\newcommand{\LLsk}  {\mathcal{L}_{\sigma_{{\rm los}, k}}}

\newcommand{\DD}    {\mathcal{D}}
\newcommand{\Rk}    {R_k}
\newcommand{\Ri}    {R_i}
\newcommand{\Rj}    {R_j}
\newcommand{\slosk} {\sigma_{{\rm los}, k}}
\newcommand{\vlosk} {v_{{\rm los}, k}}
\newcommand{\vki}   {v_{k,i}}

\newcommand{\muj}   {\mu_{{\rm V},j}}
\newcommand{\Nmu}   {N_{\mu}}
\newcommand{\Nsk}   {N_{\sigma_{{\rm los},k}}}
\newcommand{\Ns}    {N_{\sigma_{\rm los}}}
\newcommand{\Nr}    {N_R}
\newcommand{\Nvi}   {N_{v,i}}
\newcommand{\losvdki} {\mathscr{L}_{k,i}}

\newcommand{\Rinfl} {R_{\rm infl}}

\newcommand{\losvd} {\mathscr{L}}
\newcommand{\dlos}  {\Sigma_\star}
\newcommand{\slos}  {\sigma_{\rm los}}
\newcommand{\vlos}  {v_{\rm los}}

\newcommand{\dd}    {{\rm d}}
\newcommand{\dm}    {{\rm dm}}

\defcitealias{Bustamante2021}{BR21}
\defcitealias{Mateo2008}{M08}

\begin{document} 


\title{The central black hole in the dwarf spheroidal galaxy Leo I \\ Not supermassive, at most an intermediate-mass candidate }
\titlerunning{The central BH of Leo I}

\author{
R. Pascale \inst{1} \thanks{\email{raffaele.pascale@inaf.it}} \and
C. Nipoti \inst{2} \and 
F. Calura \inst{1} \and
A. Della Croce \inst{1,2}
}
\institute{
INAF - Osservatorio di Astrofisica e Scienza dello Spazio di Bologna, Via Gobetti 93/3, 40129 Bologna, Italy 
\and
Dipartimento di Fisica e Astronomia 'Augusto Righi', Università di Bologna, via Piero Gobetti 93/2, 40129 Bologna, Italy
}
\authorrunning{Pascale et al.}
\date{Received ...; accepted ...}
 
\abstract
{It has been recently claimed that a surprisingly massive black hole (BH) is present in the core of the dwarf spheroidal galaxy (dSph) Leo I. Based on integral field spectroscopy, this finding challenges the typical expectation of dSphs hosting BHs of intermediate-mass, since such a BH would better be classified as supermassive. Indeed, the analysis points toward Leo I harboring a BH with a lower mass limit exceeding a few $10^6\Msun$ at $1\sigma$, and the no BH case excluded at 95\% significance. Such a value, comparable to the entire stellar mass of the galaxy, makes Leo I a unique system that warrants further investigations. Using equilibrium models based on distribution functions (DFs) depending on actions $\fJ$ coupled with the same integral field spectroscopy data and an extensive exploration of a very large parameter space, we demonstrate, within a comprehensive Bayesian framework of model-data comparison, that the posterior on the BH mass is flat towards the low-mass end and, thus, that the kinematics of the central galaxy region only imposes an upper limit on the BH mass of few $10^5\Msun$ (at $3\sigma$). Such an upper limit brings back the putative BH of Leo I under the category of intermediate-mass BHs, and it is also in line with formation scenarios and expectations from scaling relations at the mass regime of dwarf galaxies. 
}

\keywords{
black hole physics - stars: kinematics and dynamics - methods: statistical - techniques: radial velocities 
}

\maketitle

\section{INTRODUCTION} 

The BH mass spectrum is commonly divided into three categories, depending on the BH mass ($\MBH$). Namely: stellar mass black holes (sMBHs) with $\MBH\lesssim100\Msun$, super massive black holes (SMBHs) with $\MBH\gtrsim10^5\Msun$, and intermediate-mass black holes (IMBH), comprising of everything in between \citep[$10^2\Msun\lesssim\MBH\lesssim10^5\Msun$;][]{Greene2020}. However, while sMBHs and SMBHs have been widely detected, the IMBH regime remains unpopulated, with very few detections based on gravitational waves \citep{Abbott2020} and several controversial cases \citep{Gebhardt2005,vanderMarel2010}. This observational lack is somewhat remarkable, considering that the Universe must have been, at some point, inhabited by IMBHs, as the observed SMBHs are believed to have grown from less massive ones. Additionally, the discovery of extremely massive SMBHs in young quasars at high redshifts \citep{Banados2018} poses an even greater challenge for explanation in the absence of massive seeds: how could SMBHs have grown so fast in time otherwise? 

Based on extrapolations from stellar-mass-BH mass relations, the prevailing notion is that dwarf galaxies and globular clusters (GCs) are prime locations to search for IMBHs. However, despite considerable efforts dedicated to their quest, detections remain elusive \citep{Strader2012,Lutzgendorf2016,DellaCroce2023}. In these low-mass systems, significant challenges lie, for instance, in the request for high spatial resolution to collect data from a large number of kinematic tracers close to the BH, in the accurate identification of the system's center \citep{Maccarone2005,Giersz2015,deVita2017,Pechetti2024}, or in the assumption of correct mass-to-light ratios. Nonetheless, despite the observational and theoretical obstacles, there has been a recent claim regarding the discovery of an exceptionally massive BH at the center of the local dSph Leo I \citep[hereafter BR21]{Bustamante2021}. According to this study, employing the orbit-based \cite{Schwarzschild1979} method and a frequentist model-data comparison, the $1\sigma$ BH mass is $\MBH=(3.3\pm2)\times10^6\Msun$, comparable to that of Sagittarius A$^*$. This finding is exceptional for several reasons: i) it is an actual detection, not just an upper limit, with the no BH case excluded at 95\% significance; ii) the mass of the BH is comparable with the stellar mass of the entire galaxy; iii) the BH would fall in the category of SMBHs and not of IMBHs. 

This recent discovery poses a challenge, as the BH in Leo I deviates from any scaling relation by at least two orders of magnitude \citep{Greene2020}. This exceptional deviation calls for an explanation, therefore, further investigations into Leo I are warranted to gain a deeper understanding of its unique properties. For instance, \cite{Pacucci2023} recently suggested that Leo I might have been more massive in the past, with pericentric passages close to the Milky Way (MW) potentially stripping up to 90\% of its original mass. However, while their simulations reproduce Leo I's current position and velocity dispersion, they require very strict and improbable conditions, as two pericentric passages - a scenario that is unlikely for the galaxy \citep{Sohn2013} - that are systematically close to the lower bound of the pericenter as estimated with Gaia EDR3 data \citep{Pace2022}. Nevertheless, in this letter, we tackle the problem in a different manner, and we argue that the structure and kinematics of Leo I are, instead, perfectly consistent with a no-BH scenario and that the BH, if present, is at most of intermediate mass. In our work, we fit the same dataset as in \citetalias{Bustamante2021} but using different dynamical models, based on DFs depending on actions, in a fully Bayesian framework.

\begin{figure*}
    \centering
    \includegraphics[width=1.\hsize]{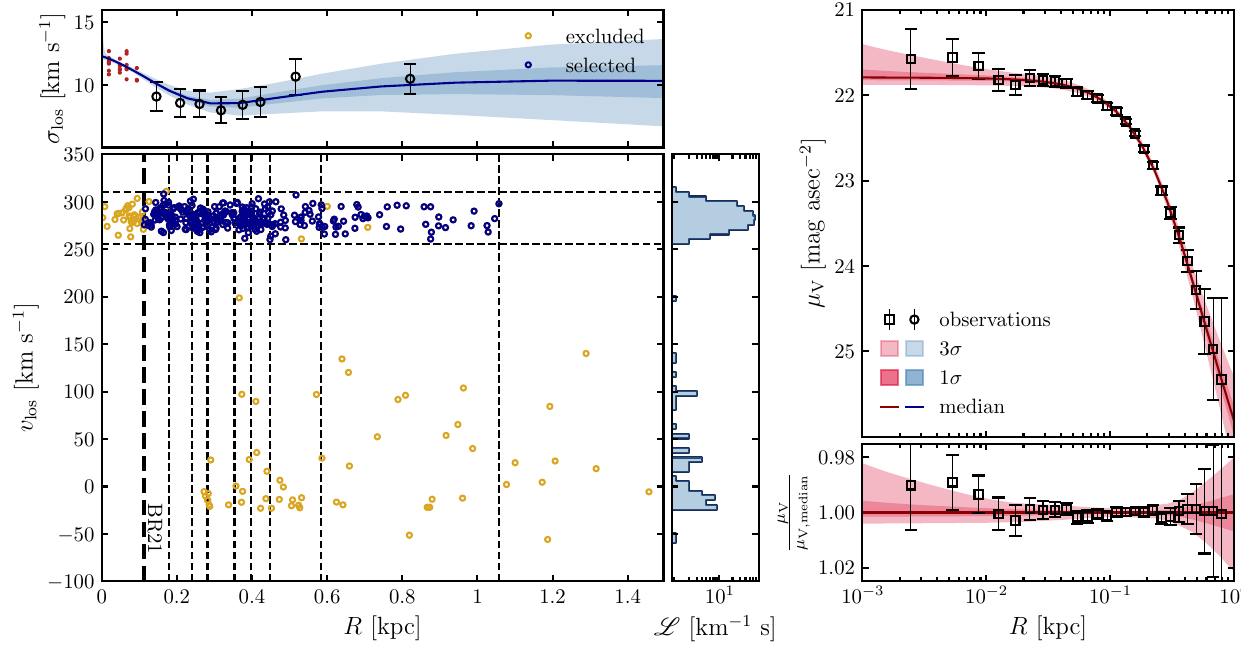}
    \caption{Bottom left: Kinematic sample from \citetalias{Mateo2008}. In blue (gold), we show stars included (excluded) in our analysis. Top left: derived los velocity dispersion profile. The bins' edges are indicated by vertical dashed lines in the bottom-left panel. The red points represent the los velocity dispersions computed from the LOSVDs shown in Fig.~\ref{fig:losvds}. Middle panel: Marginalized velocity distribution. Top right: Surface brightness profile from \citetalias{Bustamante2021}. Bottom right: Model residuals compared to the data. The light(dark) blue and red bands represent the models' 1$\sigma$ (3$\sigma$) confidence intervals, while the solid curves are median models.} 
    \label{fig:data}
\end{figure*}

\begin{figure*}
    \centering
    \includegraphics[width=1.\hsize]{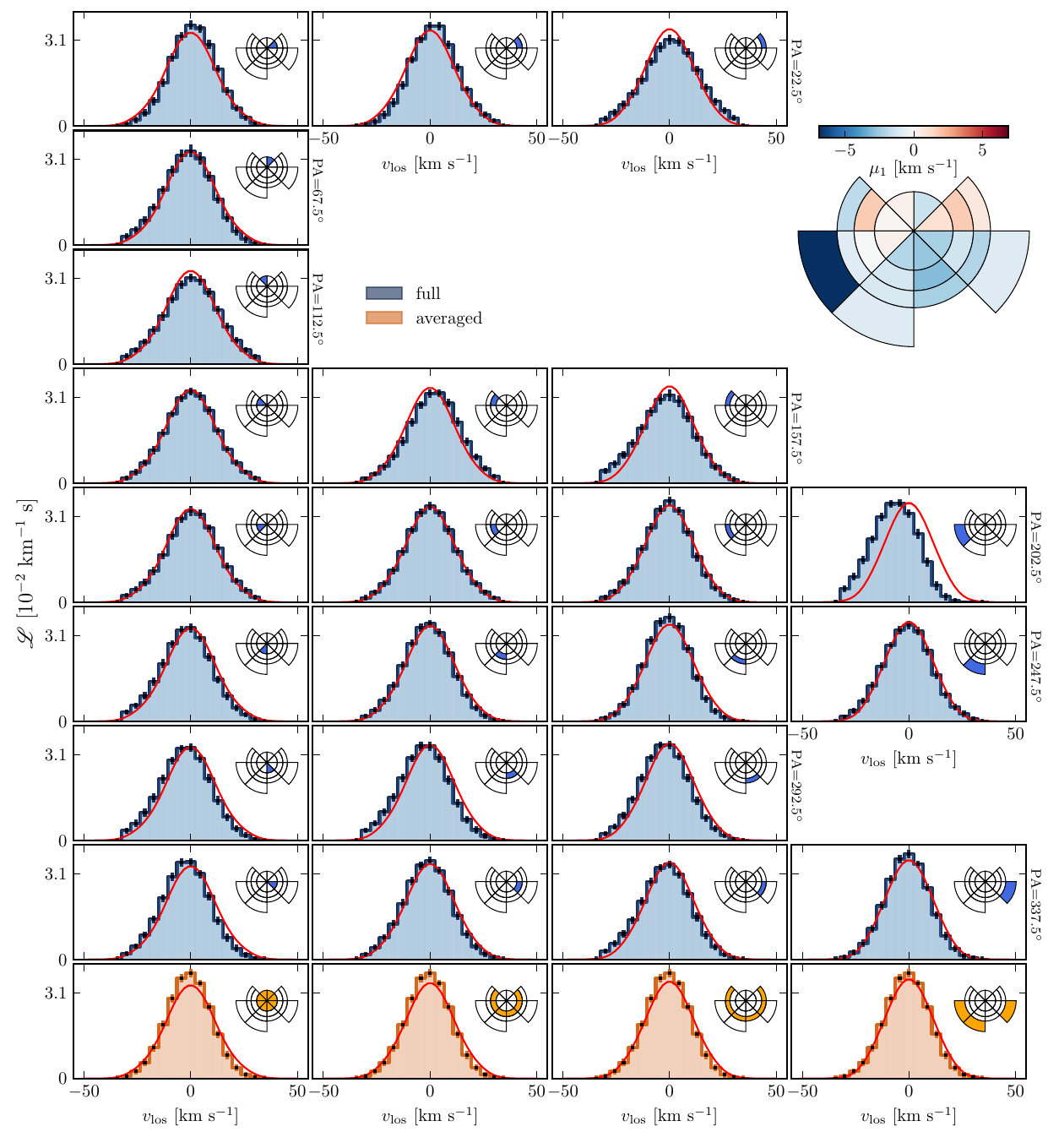}
    \caption{LOSVDs from \citetalias{Bustamante2021} (colored histograms) compared to those from the median models (red curves). The columns, from left to right, show LOSVDs computed at progressively larger radii. Rows, from top to bottom, show LOSVDs computed at increasing position angles. The bottom row (in orange) displays LOSVDs from \citetalias{Bustamante2021} weighted along the PA of each radius. The top-right part of the figure shows the spatial distribution of the LOSVDs, color-coded for the offset velocity.} 
    \label{fig:losvds}
\end{figure*}

Leo I appears as a very regular, flattened structure on the plane of the sky, with an axis ratio of $0.69\pm0.01$, similar to that of other classical dSphs \citep{Munoz2018}. It is among the brightest and most massive dSphs in the MW, with a stellar mass of $5.5\times10^6\Msun$ \citep{McConnachie2012}. Its heliocentric distance is $D=256.7\pm13.3\kpc$ \citep{Mendez2002,Bellazzini2004,Held2010,Pacucci2023}, which designates the galaxy as the most remote dSph within the MW. Throughout this work we will adopt $D=256.7\kpc$.

This letter is structured as follows: in Sections~\ref{sec:data} and~\ref{sec:models} we introduce the used dataset and  models, respectively. Section~\ref{sec:fit} describes   the fitting procedure while in Section~\ref{sec:res} we present our results and draw our conclusions. 


\section{DATA} 
\label{sec:data}

The dataset used in this work consists of the system's surface brightness profile, the central line-of-sight velocity distributions (LOSVDs) and the galaxy outer velocity dispersion profile.

\begin{itemize}
    \item {\em Photometric sample.} We adopt the surface brightness profile presented in \citetalias{Bustamante2021}, which was derived using publicly available Sloan Digital Sky Survey (SDSS) g-band imaging \citep[DR12;][]{Alam2015}. 
    This profile extends from $4\asec$ where, as pointed out by \citetalias{Bustamante2021}, it shows a mild cusp, to $20\amin$ ($\simeq1.5\kpc$ at the adopted distance). 
    The profile, shown in the right-hand panel of Fig.~\ref{fig:data}, consists, then, of $\Nmu=26$ radial bins, the $j$-th of which has average distance $\Rj$, associated V-band magnitude $\muj$ and error $\Delta\muj$. 

    \item {\em Inner kinematic sample.}
    The kinematic sample used in the central galaxy encompasses the galaxy's LOSVDs from \citetalias{Bustamante2021}\footnote{Data were privately provided by the authors.}. 
    These distributions have been computed from observations collected with the VIRUS-W spectrograph - two overlapping $1.75\times0.92\amin^2$ rectangular patches in the galaxy central regions. Due to spectrograph fiber size limitations, individual stellar spectra extraction was unfeasible, so integrated-light kinematics was employed.
    To mitigate crowding effects, the fibers were grouped into $\Nr=23$ sectors, each containing from 5 to 41 fibres, distributed across a polar grid centered on the galaxy center (see \citetalias{Bustamante2021} for details). Fig.~\ref{fig:losvds} shows the spatial distribution of the sectors (top-right), alongside with each LOSVD. Here, the $i$-th  LOSVD has a number $\Nvi$ of velocity bins and the $k$-th bin has probability $\losvdki$ and error $\Delta\losvdki$. 
    
    \item {\em Outer kinematic sample.} 
    In the galaxy's outer parts, we employ a line-of-sight (los) velocity dispersion profile based on the kinematic sample from \citet[][hereafter M08]{Mateo2008}. The profile consists of $\Ns=8$ radial bins, the $k$-th of which has average distance $\Rk$, velocity dispersion $\slosk$ and error $\Delta\slosk$.  Fig.~\ref{fig:data} shows the sample from \citetalias{Mateo2008} and the derived los velocity dispersion profile. Details on how the profile is computed are given in Appendix~\ref{app:losprof}.
\end{itemize}

\section{MODELS} 
\label{sec:models}
Leo I is represented as a multi-component galaxy with stars, dark matter (DM) and a possible BH in its center. The phase-space distribution of stars and DM is described with the DF depending on the action integrals $\JJ$
\begin{equation}\label{for:df}
\begin{split}
 \fJi & =\fni\frac{\Mi}{(2\pi\Ji)^3} \biggl[1 - \beta\frac{\Jci}{\hJi} + \biggl(\frac{\Jci}{\hJi}\biggr)^2\biggr]^{-\frac{\Gammai}{2}} \times \\
 & \biggl[1 + \biggl(\frac{\Ji}{\hJi}\biggr)^{\etai}\biggr]^{\frac{\Gammai}{\etai}} \biggl[1 + \biggl(\frac{\gJi}{\Ji}\biggr)^{\etai}\biggr]^{\frac{\Gammai-\Bi}{\etai}} \times \\
 & e^{-\bigl[\frac{\gJi}{\Jti}\bigr]^{\alphai}},
\end{split}
\end{equation}
where
\begin{equation}\label{for:hg}
\begin{split}
    & \hJi = \hri\Jr+(3-\hri-\hzi)|\Jphi|+\hzi\Jz, \\
    & \gJi = \gri\Jr+(3-\gri-\gzi)|\Jphi|+\gzi\Jz,
\end{split}
\end{equation} 
with the DM labelled by $i=\dm$ and the stars by $i=\star$ \citep{Posti2015,ColeBinney2017}. DFs depending on actions have had story of successful applications: \cite{Binney2023b, Binney2023a} applied chemo-dynamical models based on $\fJ$ DFs to model the phase-space distribution of the MW;  \cite{Pascale2018} first applied these DFs to the Fornax dSph showing that its stellar kinematics is well described only by a cored DM density distribution (see also \citealt{Pascale2019}); recently, \cite{DellaCroce2023} applied $\fJ$ models to put a very tight upper limit on the mass of a putative IMBH at the center of the GC 47 Tucanae. 

The DF (\ref{for:df}) provides the flexibility to characterize components whose density distributions are well described by a double power-law model, with the optional presence of a density core inwards, and that are truncated outwards. The velocity distribution varies with the distance from the center, shifting from one velocity bias (tangential, isotropic or radial) in the center, to another in the outer parts. The DFs (\ref{for:df}) are normalized to the target component total mass, $\Mi$, via the normalization constant $\fni$; $\Jci$ sets the size of the central component's cored region in action space and marks the transition between the cored region and the inner power law, with slope $\Gammai$. Note that when $\Jci=0$ the model has no density core. Around $\Ji$ the outer power law of index $\Bi-\Gammai$ sets in. This transition happens with a sharpness described by $\etai>0$. The DF is quickly truncated around $\simeq\Jti$, with a strength given by $\alphai\ge0$. The parameters $0<\hri<3$, $0<\hzi<3$ attribute different weights to radial and vertical orbits, thus regulating the inner velocity distribution of the models. The same task is achieved by $0<\gri<3$, $0<\gzi<3$, but in the model's outer parts. The inner and outer slopes, and the truncation of the DF in action space can be translated in similar behaviors in the physical space. 

We account for the self-gravity of both stars and DM, i.e. solving the Poisson equation
\begin{equation}\label{for:poisson}
   \nabla^2\Phii(\bbx) = 4\pi G\int \d\bbv^3\fJi,
\end{equation}
where $i\in\{\star, \dm\}$. The total gravitational potential is
\begin{equation}\label{for:phitot}
    \Phitot(\bbx) = \Phist(\bbx) + \Phidm(\bbx) + \PhiBH(\bbx),
\end{equation}
with 
\begin{equation}\label{for:BH}
    \PhiBH(r) = - \frac{G\MBH}{r}
\end{equation}
describing a third, point-like, fixed, contribution added to capture the optional presence of a central BH ($G$ is the gravitational constant). 

In the formalism of equation 
(\ref{for:hg}), the models can be flattened along the $z$-axis adjusting the weights appearing in the $\gJi$ and $\hJi$ functions. Here, however, we limit ourselves to spherical models requiring 
\begin{equation}\label{for:sph}
\begin{split}
    & 3-\hri-\hzi = \hzi, \\
    & 3-\gri-\gzi = \gzi,
\end{split}
\end{equation}
thus
\begin{equation}\label{for:gauge}
    \hzi = \frac{3 - \hri}{2},\quad\gzi = \frac{3 - \gri}{2},
\end{equation}
i.e.\ the DFs depend on actions only through the angular momentum magnitude $L=|\Jphi| + \Jz$. Without loss of generality, in the following, we will assume that $z$ is the line-of-sight and that $R$ is the distance, on the plane of the sky, from the galaxy center.

Once the DF is specified, appropriate integrations yield structural or kinematic quantities for comparison with data or predictions. Key examples include the stellar projected density
\begin{equation}\label{for:dens}
    \dlos(R) = \int\fJst\dd^3\bbv\dd z,
\end{equation}
the stellar LOSVD
\begin{equation}\label{for:losvd}
    \losvd(R,\vlos) = \frac{\int\fJst\dd^2\bbv_{\perp}\dd z}{\dlos(R)}
\end{equation}
and its second moment (i.e. the los velocity dispersion)
\begin{equation}\label{for:slos}
\begin{split}
     \slos(R) = & \int\losvd(R,\vlos)\vlos^2\dd\vlos \\ = & \frac{\int\fJst\vlos^2\dd^3\bbv\dd z}{\dlos(R)},
\end{split}
\end{equation}
with $\bbv_\perp$ the component on the plane of the sky of the velocity vector $\bbv$, perpendicular to the line-of-sight velocity $\vlos$ (where the average los velocity is null since the models are non-rotating). 


We rely on the \textsc{agama}\footnote{\url{https://github.com/GalacticDynamics-Oxford/Agama}} \citep{Vasiliev2019} software library to solve all these integrals and to compute the models' total potential. 

\section{MODEL FITTING} 
\label{sec:fit}

The log-likelihood of the model, $\ln\LL$, given the available set of data $\DD$, is
\begin{equation}\label{for:logl}
    \ln\LL = \ln\LLd + \ln\LLs + \ln\LLvd.
\end{equation}
Explicitly
\begin{equation}\label{for:loglmu}
   \ln\LLd = -\frac{1}{2}\sum_{j=1}^{\Nmu}\biggl[\frac{\muj - \log c\dlos(\Rj)}{\Delta\muj}\biggr]^2,
\end{equation}
\begin{equation}\label{for:loglslos}
   \ln\LLs = -\frac{1}{2}\sum_{k=1}^{\Ns}\biggl[\frac{\slos(\Rk)-\slosk}{\Delta\slosk}\biggr]^2,
\end{equation}
and
\begin{equation}\label{for:logllosvd}
   \ln\LLvd = -\frac{1}{2}\sum_{i=1}^{\Nr}\sum_{k=1}^{\Nvi}\biggl[\frac{\losvd(\Ri,\vki)-\losvdki}{\Delta\losvdki}\biggr]^2.
\end{equation}
Each term addresses a distinct dataset (see Section~\ref{sec:data}): equation (\ref{for:loglmu}) models the galaxy's surface brightness profile using the DF-based model (equation~\ref{for:dens}), suitably scaled into V-band surface brightness via the nuisance parameter $c$ (i.e., proportional to the V-band stellar mass-to-light ratio); equation (\ref{for:loglslos}) fits the los velocity dispersion profile, with $\slos$ derived from equation (\ref{for:slos}); equation (\ref{for:logllosvd}) represents the fit to the central LOSVDs. 

A single galaxy model is entirely determined by 17 free parameters: 
\begin{itemize}
    \item $\Mst$, $\Jst$, $\Jcst$, $\hzst$, $\gzst$, $\Gammast$, $\Bst$, $\etast$, belonging to the stellar DF~(\ref{for:df}), with $i=\star$;
    \item $\Mdm$, $\Jdm$, $\Jcdm$,  $\gzdm$, $\etadm$, $\Gammadm$, $\Bdm$ of the DM DF~(\ref{for:df}), with $i=$dm;
    \item the BH mass $\MBH$;
    \item the nuisance parameter $c$.
\end{itemize}
To decrease the number of free parameters, we also fix
\begin{itemize}
    \item $\alphast=0$ in the stellar DF since data do nor require an exponential cut-off;
    \item $\alphadm=2$ and $\Jtdm=200\kpc\kms$ in the DM DF. Here, since the outer halo density cannot be constrained by the available dataset it is meaningless to fit the halo truncation radius. By truncating the halo, we also ensure that the total DM mass is a well defined quantity. For spherical models, the exact values of $\Jtdm$ and $\alphadm$ do not matter as long as the halo is truncated beyond the radial extent of the stellar component;
    \item $\hzdm=\gzdm$ since the inner/outer anisotropy of the DM halo is an unconstrained quantity in the present dataset.
\end{itemize}

We run a Markov-Chain Monte Carlo (MCMC) procedure to sample from the posterior distribution. Further details are given in Appendix~\ref{sec:pdis}, while the prior, $1\sigma$ and $3\sigma$ confidence intervals on the models' parameters resulting from the analysis are listed in Table~\ref{tab:params}.

\begin{figure}
    \centering
    \includegraphics[width=1\hsize]{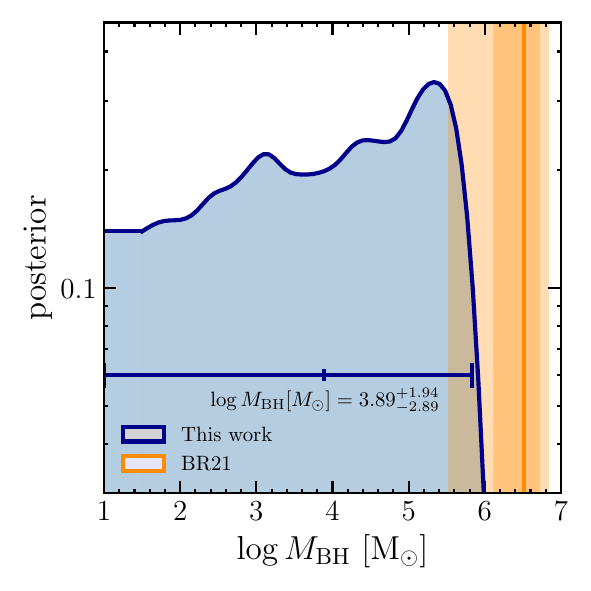}
    \caption{Posterior distribution on the logarithm of the BH mass $\log\MBH$ (represented by the blue curve). The blue horizontal bar show the $3\sigma$ confidence interval. The posterior is compared with the best model, 1$\sigma$ and $3\sigma$ estimates from \citetalias{Bustamante2021} (shown by the solid orange line and bands).  }\label{fig:posterior} 
\end{figure}

\section{RESULTS AND CONCLUSIONS} 
\label{sec:res}

Here, we present and discuss the key results of our study. Fig.~\ref{fig:posterior} shows the one-dimensional, marginalized, posterior distribution on $\log\MBH$. Over the explored prior (see Table~\ref{tab:params}), the posterior is almost uniform towards low masses, it slightly increases around $\log\MBH\simeq5.5$
whereas it has a sharp truncation around $\sim10^6\Msun$. This strongly suggests that there is insufficient statistical evidence to assert the detection of a BH. Our inference is limited at establishing an upper limit on the BH mass, set to $6.76\times10^5\Msun$ at $3\sigma$. Our $3\sigma$ upper limit is lower than the $1\sigma$ lower limit ($\simeq1.3\times10^6\Msun$) reported by \citetalias{Bustamante2021}, but, as
it can be appreciated from Fig.~\ref{fig:posterior}, is marginally consistent with the $3\sigma$ lower bound derived from the reference model of \citetalias{Bustamante2021} (vertical band in Fig.~\ref{fig:posterior}). We note that, beside their reference model, \citetalias{Bustamante2021} present other 3 models for all of which they infer a $1\sigma$  lower limit on $\MBH$ larger than $10^6\Msun$. However, upon careful examination of these models, it turns out that they are consistent with a no-BH scenario at $3\sigma$ level. We stress that \citetalias{Bustamante2021} do not report explicitly the $3\sigma$ intervals on their models' free parameters, so we computed them directly from their Figure 11, and in particular from the panels plotting $\chi^2$ as a function of the BH mass. Consistent with the frequentist approach used in \citetalias{Bustamante2021}, we computed the $3\sigma$ interval on the BH mass as the interval with $\Delta\chi^2=3$ with $\Delta\chi^2$ the $\chi^2$ difference from the minimum $\chi^2$ - the best fit model.

When compared to the spatial extent of the LOSVDs dataset, the inferred BH mass translates into an upper limit on the radius of influence $\Rinfl\equiv G\MBH/\sigma_{\rm los}$ ($\sigma_{\rm los}$ is los velocity dispersion) of the BH that is comparable to the average distance of the innermost kinematic sectors ($\simeq15.35\asec=19\pc$). In the context of \citetalias{Bustamante2021}'s study, instead, the inferred BH mass implies a well defined $\Rinfl$, comparable to the distance of the outer LOSVDs sectors. Therefore, in our case, the kinematics of the inner regions effectively exclude the existence of a SMBH rather than supporting its presence.

In Fig.s~\ref{fig:data} and \ref{fig:losvds}, we provide an illustration of the agreement between data and models. Fig.~\ref{fig:data} shows the models' predictions for the los velocity dispersion and the surface brightness profiles compared to observations. Fig.~\ref{fig:losvds} depict the LOSVDs from \citetalias{Bustamante2021} and the analogous from the models. In both cases, the models accurately fit the data.

Based on scaling relations \citep[e.g.][Figure 3]{Greene2020}, it is inferred that for a galaxy similar to Leo I, with a velocity dispersion between 10-12$\kms$, one would expect an IMBH of, at most, $10^4\Msun$, or equivalently $\mu\equiv{\MBH}/{\Mst}=10^{-4}-10^{-3}$. Our upper limits on the BH mass and $\mu$ are in well agreement with these values. In terms of $\mu$, we measure a median $\mu=0.0013$, with a $3\sigma$ upper limit at 0.033. As previously explained, these values represent the lower bounds that can be imposed by the kinematic dataset of LOSVDs.

The use of integral field spectroscopy to look for IMBH in dense stellar systems has been extensively debated in the past. In GCs, it is believed, for instance, that the methodology may introduce biases, as the collected spectra can be dominated by a few bright stars, rather than sampling the underlying stellar distribution. It is emblematic the case of NGC 6388 where spectra from individual stars sample a central velocity dispersion of $\simeq10\kms$ \citep{Lanzoni2013}, while values from integral field spectroscopy indicate a dispersion as high as $25\kms$ \citep{Lutzgendorf2011}, which end up being interpreted as signatures of massive BHs. The case of Leo I is undoubtedly more enigmatic and it is highly challenging to conclusively pinpoint the reason behind the discrepancy between our results and those from \citetalias{Bustamante2021}. In contrast to NGC 6388 where two datasets yield different results, we employ the same dataset of LOSVDs as in \citetalias{Bustamante2021}, but with a different fitting algorithm and models. Thus, the use of integral field spectroscopy cannot be the reason behind these differences.

We believe that the differences in the results stem from an interplay of factors. As pointed out by \citetalias{Bustamante2021} (see also Fig.~\ref{fig:data}), the central los velocity dispersion of Leo I inferred from the LOSVDs rises showing a mild cusp with respect to the central los velocity than can be measured from individual star velocities, aligning with the findings of \cite{Lanzoni2013}. While the application of integral field spectroscopy introduces a potential bias towards favoring the presence of a massive BH, the integration of more generalized models allows for improved marginalization over key stellar properties, as stellar anisotropy, or stellar and DM distributions. These models effectively mitigate the initial bias, thereby contributing to a more nuanced interpretation of the findings. A potential resolution to this issue may be using individual star velocities sampling the region given by the influence radius of the BH. Nevertheless, the considerable distance of Leo I poses a significant challenge. Assuming the LOSVD shape is crucial for detecting a BH, \cite{AmoriscoEvans2012} estimate that a sample of at least 100 objects with an error smaller than 0.2$\slos$ is required to recover any velocity distribution. In the case of Leo I, this implies more than 100 stars with a $2\kms$ error in the los velocity, confined in a region smaller than 15 arcseconds, in order to infer a BH with a mass a few $10^5 \Msun$, as constrained by our upper limit.

\bibliographystyle{aa}
\bibliography{main}



\begin{table*}[h]
\centering
\begin{tabular}{lccccc}
\hline
Parameter & Prior & median & $1\sigma$ & $3\sigma$ \\
\hline
$\log\Mst$ [$\Msun$]        & [6.439, 7.041]      & 6.759 & [6.567, 6.939] & [6.441, 7.040] \\
$\log\Jcst$ [$\kpc\kms$]    & [-2, 0]  & -0.378 & [-0.453, -0.316] & [-0.698, -0.190] \\
$\log\Jst$ [$\kpc\kms$]     & [0, 1]   & 0.460 & [0.441, 0.479] & [0.401, 0.522] \\
$\hzst$                     & [0, 1.5] & 0.783 & [0.737, 0.825] & [0.639, 0.929] \\
$\gzst$                     & [0, 1.5] & 1.024 & [1.008, 1.040] & [0.977, 1.069] \\
$\Gammast$                  & [0, 3]   & 1.378 & [1.289, 1.461] & [1.040, 1.616] \\
$\Bst$                      & [3, 9]   & 6.576 & [6.336, 6.805] & [5.918, 7.295] \\
$\etast$                    & [0.5, 10]& 7.377 & [5.489, 8.979] & [3.315, 9.985] \\
$\log\Mdm$ [$\Msun$]        & [6, 11]  & 8.846 & [8.561, 9.053] & [8.091, 9.422] \\
$\log\Jcdm$ [$\kpc\kms$]    & [-1.5, 0]& -0.394 & [-1.089, 0.320] & [-1.497, 1.128] \\
$\log\Jdm$ [$\kpc\kms$]     & [0, 2.5] & 1.707 & [1.321, 2.170] & [0.335, 2.494] \\
$\hzdm$                     & [0, 1.5] & 0.927 & [0.557, 1.227] & [0.130, 1.416] \\
$\Gammadm$                  & [0, 3]   & 1.301 & [0.974, 1.567] & [0.146, 2.156] \\
$\Bdm$                      & [3, 10]  & 5.479 & [3.854, 7.803] & [3.011, 9.983] \\
$\etadm$                    & [0.5, 5] & 2.985 & [1.493, 4.310] & [0.522, 4.996] \\
$\log\MBH$ [$\Msun$]        & [1, 7]   & 3.894 & [2.105, 5.274] & [1.005, 5.835] \\
$c$                         & [25, 35] & 29.773 & [29.581, 29.952] & [29.449, 30.063] \\
\hline
\end{tabular}
\caption{List of free parameters alongside the priors adopted, the median values, $1\sigma$ and $3\sigma$ confidence intervals resulting from the analysis.}\label{tab:params}
\end{table*}

\appendix

\section{DERIVATION OF THE LOS VELOCITY DISPERISON PROFILE}
\label{app:losprof}
The sample of radial velocities from \citetalias{Mateo2008} used to compute the galaxy's los velocity disperion profile consists of 387 stars in the region of the sky occupied by Leo I. They have been obtained with the Hectochelle multiobject echelle spectrograph on the MMT. To keep our dataset as similar as possible to that of \citetalias{Bustamante2021}, we removed all stars inside a $90.3\asec$ radius, corresponding to the radial extent of the outer LOSVD sectors. Also, we removed all stars flagged as possible binaries or false positives by \citetalias{Mateo2008}, i.e.\ observations in which the velocity of the sky could have been measured instead of the target star velocity (see \citetalias{Mateo2008} for details). We performed an iterative $3\sigma$ clip on the remaining stars to remove further foreground stars. As discussed by \citetalias{Mateo2008} the distinction between members and foreground stars is very clear, i.e. the systemic velocity of Leo I is very different from the average velocity of foreground stars in that direction ($\Delta v\simeq 250\kms$). The final sample consists of 288 stars with median error of $2\kms$. 

Using these stars, we construct the galaxy's los velocity dispersion profile. Each radial bin contains 35 stars, except the last bin that contains 43 stars. We adopt the method of \cite{Pryor93} for computing the galaxy velocity dispersion. The likelihood of having velocity dispersion $\slosk$ in the $k$-th radial bin reads
\begin{equation}\label{for:pryor}
        \LLsk = \prod_{i=1}^{\Nsk}\frac{e^{-\frac{(\vki-\vlosk)^2}{2(\slosk^2 + \delta\vki^2)}}}{\sqrt{2\pi(\slosk^2 + \delta\vki^2)}}.
\end{equation}
In the above equation, $\vki$ and $\delta\vki$ are the $i$-th radial velocity and corresponding error, $\slosk$ and $\vlosk$ are the velocity dispersion and average velocity of the bin (i.e.\ the models' free parameters), while the sum extends over the stars falling in the bin ($\Nsk=35$ or 43). We run an MCMC procedure very similar to that of Appendix~\ref{sec:pdis} to sample from the posterior. We take as a measure of the velocity dispersion of the $k$-th bin the median value of the marginalized, one-dimensional, posterior distribution and, as error, the average distance between the 84th-50th and 50th-16th percentiles. 

\section{MCMC FITTING PROCEDURE}
\label{sec:pdis}

In the MCMC fit of Section~\ref{sec:fit}, to sample from the posterior we used 150 chains, each evolved for 20000 steps approximately. We combine the differential evolution proposal by \cite{Nelson2014} and the snooker proposal by \cite{terBraak2008}, as implemented in the software library \textsc{emcee} \citep{ForemanMackey2013}. %
During each MCMC iteration, we randomly select one of two proposals assigning a 60\% probability to the differential evolution proposal with a stretch factor of $0.8\times({2.38}/{\sqrt{2n}})$, where $n=17$ (the dimension of the parameter space), and a 40\% probability to the snooker proposal with a 0.85 stretch factor. Empirically, this combination improves MCMC performance by reducing chain autocorrelation and increasing acceptance rates.

\begin{figure*}
    \centering
    \includegraphics[angle=270,width=1\hsize]{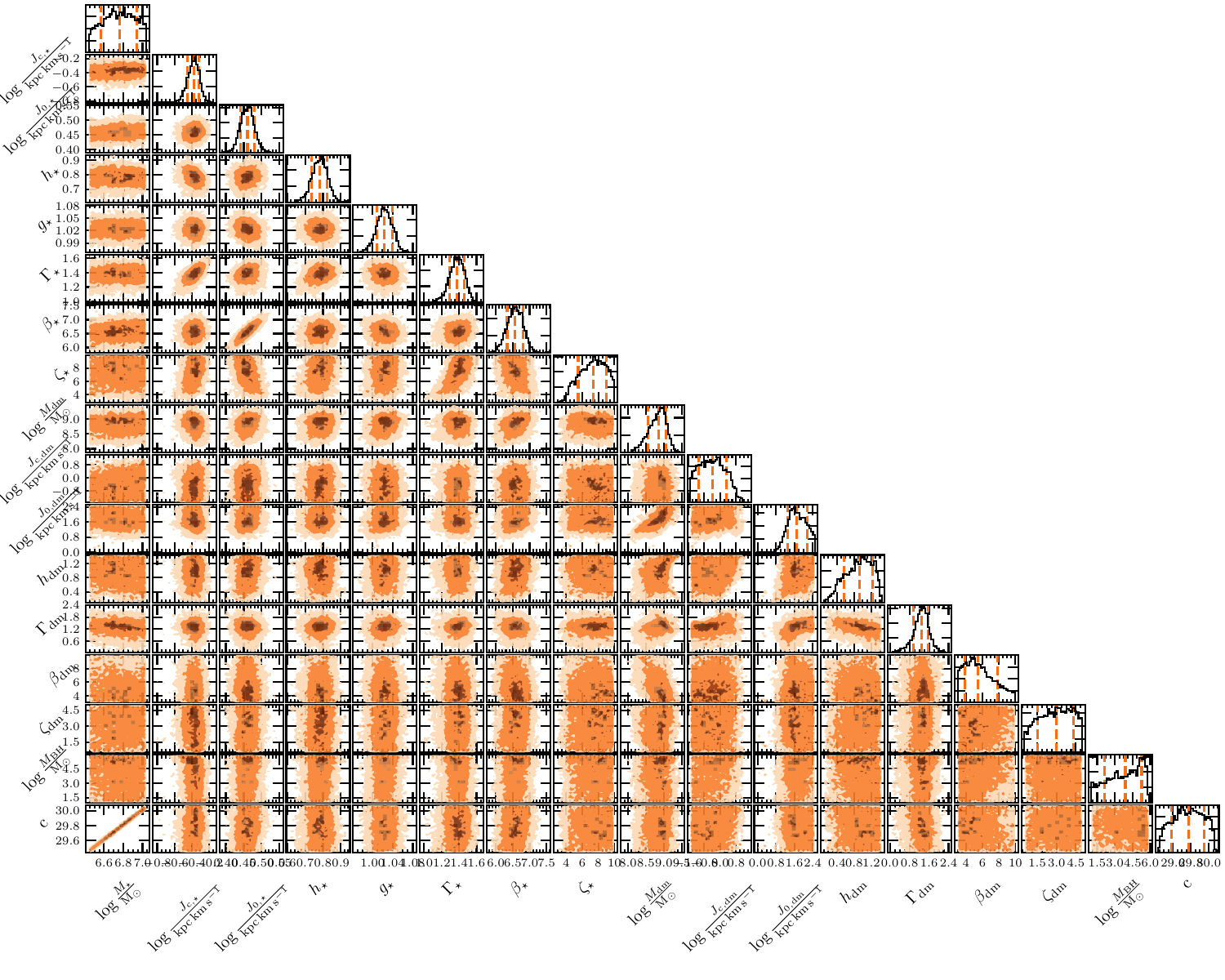}
    \caption{Two- and one-dimensional marginalized posterior distributions over the model free parameters (see Table~\ref{tab:params}). The dashed-orange curves in the one-dimensional posterior distributions mark the 16th, 50th and 84th percentiles, used to evaluate the $1\sigma$ confidence intervals on the corresponding parameters.}
    \label{fig:corner}
\end{figure*}

Fig.~\ref{fig:corner} shows the one- and two-dimensional posterior distributions over the models' free parameters. A zoom-in view of the posterior distribution over $\log\MBH$ is shown in Fig.~\ref{fig:posterior}. All walkers except for one, that did not reach convergence, have been used to build the posterior and evaluate all confidence intervals. The initial 5000 steps from each walker were discarded as burn-in, ensuring low autocorrelation and model convergence. We have further applied a thinning of 50 steps, approximately of the order of the chains' autocorrelations length. The $1\sigma$ confidence intervals over the models parameters or any derived quantity are estimated as the 16th, 50th and 84th percentiles of the corresponding distributions, while the $3\sigma$ confidence intervals as the 0.15th, 50th and 99.85th percentiles.


\begin{acknowledgements}
We thank the anonymous referee for the useful comments and suggestions, which helped improving the quality of the work. We are very thankful to Maria Jose Bustamante-Rosell for sharing data. 
This paper is supported by the Italian Research Center on High Performance Computing Big Data and Quantum
Computing (ICSC), project funded by European Union - NextGenerationEU - and National Recovery and Resilience Plan (NRRP) - Mission 4 Component 2 within the activities of Spoke 3 (Astrophysics and Cosmos Observations). 
The research activities described in this paper have been co-funded by the European Union - NextGeneration EU within PRIN 2022 project n.20229YBSAN - Globular clusters in cosmological simulations and in lensed fields: from their birth to the present epoch.
\end{acknowledgements}

\end{document}